\documentstyle[prl,aps,multicol]{revtex}
\begin{document}
\draft
\widetext

\newcommand{\be}{\begin{equation}}
\newcommand{\ee}{\end{equation}}
\newcommand{\ber}{\begin{eqnarray}}
\newcommand{\eer}{\end{eqnarray}}

\title{Universal macroscopic background formation in surface super-roughening}

\author{H.-W. Lee$^{(1)}$ and Doochul Kim$^{(2)}$}
\address{
$^{(1)}$Center for Theoretical Physics, Seoul National University,
Seoul 151-742, Korea \\
$^{(2)}$Department of Physics and Center for Theoretical Physics,
Seoul National University, Seoul 151-742, Korea}

\maketitle

\widetext
\begin{abstract}
We study a class of super-rough growth models whose
structure factor satisfies the Family-Vicsek scaling.
We demonstrate that a macroscopic background spontaneously
develops in the local surface profile, which dominates
the scaling of the local surface width and the height-difference.
The shape of the macroscopic background takes a form of 
a finite-order polynomial whose order 
is decided from the value of the global roughness exponent. 
Once the macroscopic background
is subtracted, the width of the resulting local surface profile 
satisfies the Family-Vicsek scaling. 
We show that this feature is universal to all super-rough growth
models, and we also discuss the difference
between the macroscopic background formation and the pattern formation
in other models.
\end{abstract}
\pacs{PACS number(s): 05.40.+j, 05.70.Ln, 68.35.Fx }


\widetext
 
Recently, there has been considerable progress in the understanding
of the dynamics of growing surfaces \cite{Family,Barabasi}. 
Much of the progress
is motivated by the crucial observation that surface roughening
exhibits scaling behaviors.
For example, the growth of the global surface width $w(L,t)$,
when it starts from the flat surface, shows the following
behavior. 
At an initial stage, $w(L,t)$ grows as a power law of time 
$w(L,t) \sim t^{\beta}$, where $\beta$ is called the growth
exponent. At a later stage, however, 
$w(L,t)$ saturates 
to a certain power of the system size $L$, $w(L,t) \sim L^{\alpha}$, 
where $\alpha$ is called the roughness exponent.
The crossover to the saturated surface is governed by
the lateral correlation length $\xi(t)$, which
scales as $t^{1/z}$ for the initial stage $t \ll L^z$ and
saturates to $L$ for the later stage $t \gg L^z$.
Here the third exponent $z$ is called the dynamic exponent, and
this scaling behavior is called the Family-Vicsek (FV) scaling 
ansatz \cite{FVansatz}.

The presence of the scaling is usually related to the self-affine 
structure of a surface, implying the same properties
at all length scales.
In some growth models (for example, Refs.~\cite{Amar,Amaral}), however, 
surface properties exhibit
different behaviors when they are probed at different length
scales. For example, one may probe the surface width $w(l,t)$
within a local window of size $l$ $(\ll L)$, where
\be
w^2(l,t)=\{ \langle 
( h(x,t)-\langle h(x,t) \rangle_{X_l}
)^2 \rangle_{X_l} \} \ , 
\label{eq:w2def}
\ee
and $\langle \ \rangle_{X_l}$ is the spatial
average over the local window $X_l$ of size $l$
and $\{ \ \}$ the sample average.
And one finds
\be
w^2(l,t) \propto \left\{
\begin{array}{ll} \displaystyle
t^{2\beta} & \mbox{for } t\ll l^z \\
l^{2\alpha}
  \displaystyle\left( \xi(t) \over l \right)^{2(\alpha-\alpha')}
   & \mbox{for } t \gg l^z
\end{array} \right. .
\label{eq:AR}
\ee
Note that while it follows the FV scaling at the
initial stage $t\ll l^z$,
the scaling behavior at the later stage is {\it anomalous} :
the window size dependence is described not by the {\it global} 
roughness exponent $\alpha$, but rather by a new exponent 
$\alpha'(<\alpha)$ which is called {\it local} roughness exponent. 
Note also that at the later stage, 
the surface width implied by Eq.~(\ref{eq:AR}) is
anomalously larger than the FV scaling result $l^{2\alpha}$,
since $\alpha>\alpha'$ and $\xi(t) \gg l$.

In the past few years, there have been many works on
the anomalous roughening. Numerical simulations demonstrated 
the occurrence of the anomalous roughening in some growth models
(for example, see Refs.~\cite{Amar,Amaral}).
New scaling ansatz for the anomalous roughening was
proposed \cite{Schroeder,DasSarma}. Very recently,
L\'{o}pez {\it et al.} \cite{Lopez1997} identified two 
separate mechanisms of the anomalous roughening, super-roughening
and intrinsic roughening. In case of the super-roughening $(\alpha>1)$,
the structure factor $S(k)$, or the power spectrum, follows
the Family-Vicsek scaling and the anomalous roughening occurs due to
the divergence of $S(k)$ near $k=0$.
In case of the intrinsic anomalous roughening, on the other hand,
$S(k)$ itself scales anomalously.

Among the two mechanisms, we focus in this paper
on the anomalous roughening due to the super-roughening. 
In particular, we study
the morphology of the local surface profile generated in
the super-rough growth models. Usually, the surface profile
can be probed from the scaling of the local surface width or
the height-difference as a function of a probing length scale.
In super-rough surfaces, however, the scaling is much less
informative, since the relevant local roughness exponent $\alpha'$
is always 1 \cite{Lopez1997} independent of many details of growth
models, the only implication of the universal value 1 being the
divergence of $S(k)$.  
So to probe the surface profile in the super-rough growth models,
we take a different theoretical approach based on 
the least-square-fitting method. 
As a main result of the paper, we find that
(i) the local surface profile forms a macroscopic background that takes
the form of a finite order polynomial and that
(ii) the magnitude of the short wavelength fluctuations superimposed
on the macroscopic background follows the FV scaling.
We also show that this property is universal to all super-rough
growth models.

One simple way to study the profile is to make a guess on 
the functional form of the profile, fit the function to
the profile through the least-square-fitting method,
and examine the magnitude of the discrepancy. 
The simplest example of this approach is the local surface width
since the idea of Eq.~(\ref{eq:w2def}) is nothing
but the least-square-fitting of the local surface
profile with a constant function $\langle h(x,t) \rangle_{X_l}$.
Adopting this view point, it is then interesting to
explore other possibilities of fitting functions,
which hopefully produces a smaller discrepancy. 
Since the origin of the anomalous roughening in the super-rough
growth models is the divergence in the long wavelength components
that appear smooth in the local window,
we first take a first-order polynomial in $x$ as a fitting function : 
$\tilde{h}_1(x,t)=a_1(t)x+a_0(t)$.
Here, the coefficients $a_i(t)$ are fixed by minimizing the discrepancy
$\langle [h(x,t)-\tilde{h}_1 (x,t)]^2 \rangle_{X_l}$
for a given surface height profile $h(x,t)$.

For illustration, we first use the 1+1 dimensional uniform 
diffusion models as examples :
\be
{\partial h(x,t) \over \partial t}=
(-1)^{m+1}{\partial^{2m} h(x,t) \over \partial x^{2m}} + \eta (x,t) ,
\label{eq:MBE}
\ee
where $\eta (x,t)$ is the noise with correlation
$\{ \eta(x,t) \eta (x',t') \}$ $=$$D \delta (x-x')\delta(t-t')$,
and $m=1,2,3, \ldots \ $.
For an arbitrary $m$, it can be explicitly verified that 
the structure factor follows the FV scaling.
Also the precise values of the scaling exponents are known :
$\alpha=(2m-1)/2, \beta=(2m-1)/4m$,
and $z=2m$ \cite {Lopez}.
Notice that for $m \ge 2$, the uniform diffusion models become super-rough 
($\alpha>1$). Then as
demonstrated in Ref.~\cite{Lopez1997}, the models exhibit 
the anomalous scaling with $\alpha'=1$ for $m \ge 2$.

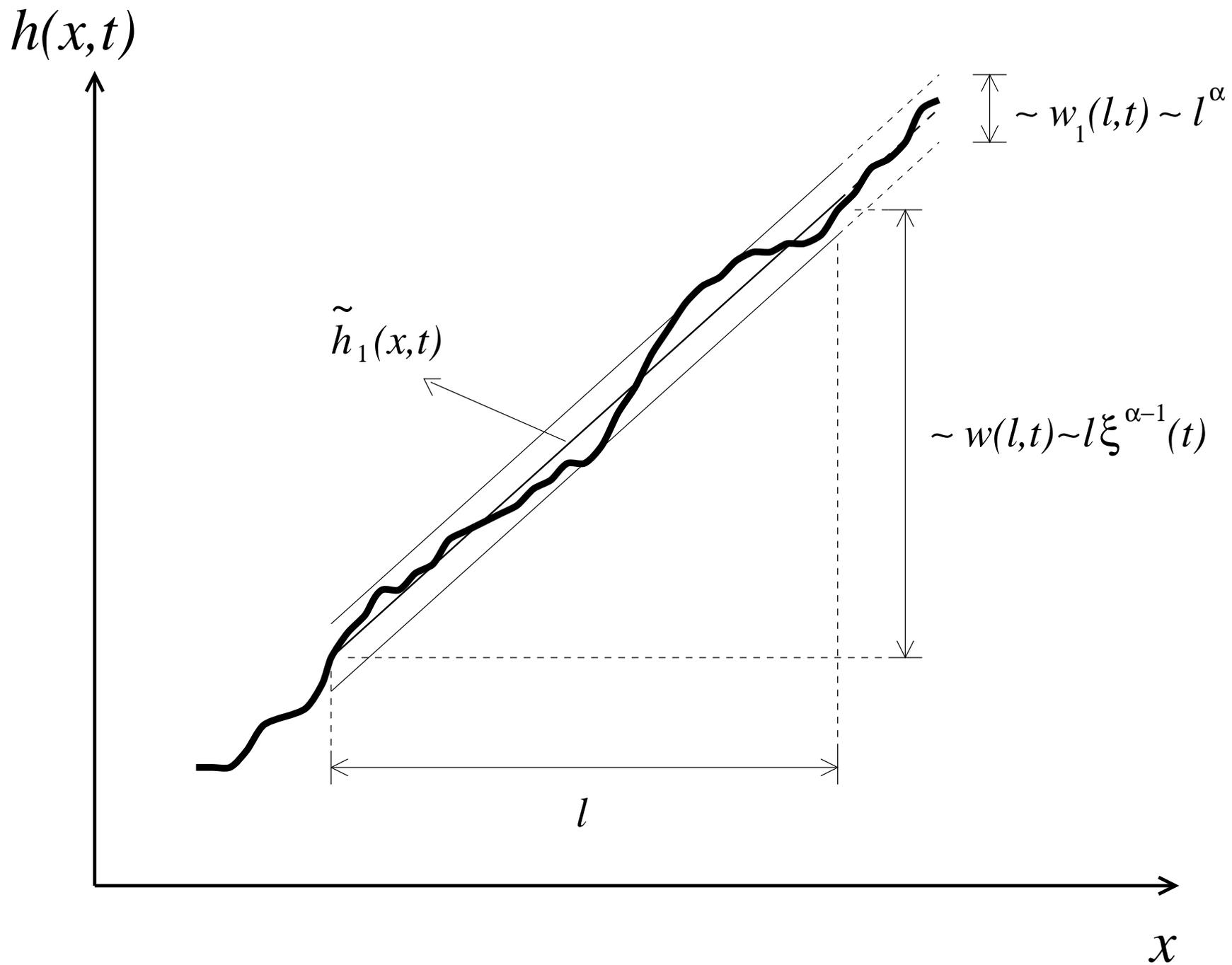
\begin{figure}

\caption{Typical surface profile of a 1+1 dimensional uniform
  diffusion model with $m=2$ [Eq.~(\protect\ref{eq:MBE})]
  at the later stage $\xi(t) \gg l$. 
  Notice the formation of a linear macroscopic background 
  $\tilde{h}_1(x,t)$ whose large slope leads to the anomalous
  scaling of $w^2(l,t)$.
  Once the background is subtracted, 
  the resulting local surface width $w_1(l,t)$ is characterized 
  by the {\it global} roughness exponent $\alpha$ and satisfies
  the Family-Vicsek scaling.
}
\label{fig:groove}
\end{figure}

The accuracy of the fitting with the first-order polynomial
can be estimated from the ensemble averaged minimal discrepancy,
\be
w^2_1(l,t)=\{ \langle ( h(x,t)-\tilde{h_1}(x,t) 
  )^2 \rangle_{X_l} \} \ ,
\label{eq:modw21orig}
\ee 
where we call $w_1(l,t)$ first-order generalized surface width.
For the linear models Eq.~(\ref{eq:MBE}),
$w_1(l,t)$ can be evaluated analytically in a straightforward way.
At the initial stage $\xi(t) \ll l$, $w^2_1(l,t)$ scales
the same way as $w^2(l,t)$, implying that there is no substantial
improvement of the fitting from the new choice of 
the fitting function. This result is understandable since
the divergence of $S(k)$ does not qualitatively affect 
the surface profile at the initial stage.
At the later stage $\xi(t) \gg l$,
however, the results are 
\be
w^2_1(l,t) \propto \left\{
  \begin{array}{ll}
  l^{2\alpha} & \mbox{for } \alpha<2 \mbox{  (or } m=1,2) \  \\
  l^{2\alpha} 
    \displaystyle\left( {\xi(t) \over l} \right)^{2(\alpha-2)} 
     & \mbox{for } \alpha >2 \mbox{  (or } m \ge 3). 
  \end{array} \right.
\label{eq:w21MBE}
\ee 
Here the reduction in  the power of the large factor $\xi(t)/l$
should be noticed : it has been reduced from 
$2(\alpha-1)$ in $w^2(l,t)$ to
$0$ for $m=2$ and to $2(\alpha-2)$ for $m \ge 3$.
For the definiteness, let us take $l/\xi(t)$ 
as a small parameter of the analysis. Then the power reduction
implies the substantial improvement of the fitting,
which we interpret as an evidence of a spontaneous formation 
of a {\it macroscopic background} in the surface profile.
Here the word ``macroscopic'' denotes that the characteristic length
scale of the background is comparable to or larger than the window size $l$.
Figure~\ref{fig:groove} shows a typical surface profile of 
the $m=2$ uniform diffusion model at the later stage $\xi(t) \gg l$, 
and the formation of the linear background is clear \cite{commentAmar}.

Motivated by the success of the first-order polynomial,
we explore the idea further and examine the other types of fitting
functions. Specifically we choose a higher-order polynomial, which is a
natural extension of the first-order polynomial. One merit of this
choice is that polynomials form a complete set of basis for functional
space. So by examining
the fitting with high-order polynomials, one can examine the effect of 
all types of fitting functions. 
To estimate the accuracy of the fitting with
an $N$th order polynomial, we introduce
an $N$th order generalized surface width $w_N(l,t)$, where
\be
w^2_N(l,t)=\{ \langle [ h(x,t)-\tilde{h_N}(x,t) 
  ]^2 \rangle_{X_l} \} \ ,
\label{eq:modw2Norig}
\ee
and the coefficients $a_i(t)$ of 
$\tilde{h}_N(x,t)=\sum_{i=0}^{N} a_i(t)x^i$ are similarly
fixed by the least-square-fitting method for a given realization $h(x,t)$.
For the linear model Eq.~(\ref{eq:MBE}), 
one can analytically verify that $w^2_N(l,t)$ scales the same
way as $w^2(l,t)$ at the initial stage, again finding no improvement, 
and 
\be
w^2_N(l,t) \propto \left\{
  \begin{array}{ll}
  l^{2\alpha} & 
  \mbox{for } \alpha<N+1 
    \mbox{ (or } m \le N+1) 
  \  \\
  l^{2\alpha} 
    \displaystyle\left( {\xi(t) \over l} \right)^{2(\alpha-N-1)} & 
  \mbox{for } \alpha >N+1  
     \mbox{ (or } m \ge N+2)
  \end{array}
 \right. \ ,
\label{eq:w2Nscaling}
\ee 
at the later stage.
This scaling result contains very interesting information.
First,
the accuracy improves substantially as the order $N$ increases 
till $N$ becomes larger than $\alpha-1$,
after which it saturates and does not improve
upon the further increase of $N$.
Recalling the completeness of the polynomial basis,
one then realizes that even if other forms of fitting function were chosen,
such as sines and cosines, they would not improve the accuracy either. 
In this sense, one can say that the macroscopic background in
the uniform diffusion models takes the form of 
an $N$th order polynomial, where $N$ is the largest integer smaller 
than $\alpha$.
Second,
the minimal discrepancy of the fitting is always of the order
of $l^{2\alpha}$ \cite{commentN}, and so it is characterized 
by the {\it global} roughness exponent. 
One interesting consequence of this result is that
once the macroscopic background is subtracted, the width of 
the resulting local surface profile satisfies the FV scaling. 

To understand the underlying physics in the uniform diffusion models, 
let us represent the local surface
profile in the following way : $h(x,t)=\sum_{i=0}^{2m-1} a_i(t) x^i
+\sum_{k}[ A(k,t) \sin kx + B(k,t) \cos kx]$, where $kl$ is an integer 
multiple of $2\pi$.
One interesting property of the uniform diffusion
models is that while the relaxation force tends to suppress
the amplitude of the sinusoidal part, 
it does not affect the polynomial part : 
$F[\sum_{i=0}^{2m-1} a_i(t) x^i]=0$ for arbitrary $a_i(t)$ 
where $F[h(x,t)]=(-1)^{m+1} \partial^{2m}h/\partial x^{2m}$.
This property opens a possibility for the anomalous growth of
the coefficients $a_i(t)$, which makes polynomials special candidates
for a fitting function.
And the scaling result of $w^2_N(l,t)$ implies that 
the coefficients $a_i(t)$ with $i<\alpha$ (or $i<m$) 
indeed become anomalously large so that they lead to the formation of 
the macroscopic background.

Let us compare the formation of the macroscopic background
in the uniform diffusion models with pattern formation 
in other models where the pattern formation is due to
symmetry breaking or the coherent superposition of modes
of different wavelength \cite{pattern}. In the uniform diffusion
models, the symmetry breaking does not occur and the ensemble 
average of $a_i(t)$ vanishes. Also there is no coherent coupling 
between different modes since the models are linear.
Here, however, the fluctuations of $a_i(t)$ in different realizations 
are so large that the fluctuations lead to the macroscopic
background formation for any given realization,
that is, the macroscopic background formation is due to
large {\it fluctuations} in some particular ``degrees of freedom''. 
In this sense, the macroscopic background formation
is completely different from the pattern formation in its origin.

For the 1+1 dimensional uniform diffusion models,
it is now demonstrated that (i) a macroscopic background develops
spontaneously in the local surface profile, and the background
takes a form of an $N$th order polynomial where $N$ is the largest
integer smaller than $\alpha$, 
and that (ii) once the macroscopic background is subtracted,
the resulting local surface profile satisfies the FV scaling.
Below, we show that these properties are universal to
all super-rough growth models.

From now on, we restrict the discussion to the later stage
since it is trivial to show that $w^2_N(l,t)$ satisfies
the FV scaling at the initial stage for 
an arbitrary $N$.
To study the scaling behavior at the later stage, it is useful
to first relate the height-height correlation
function $G(x,t)$
to the structure factor $S(k,t)=\{ \hat{h}(k,t)\hat{h}(-k,t) \}$.
By comparing the definitions, one can obtain the following
relation between $G(x,t)$ and $S(k,t)$ \cite{Lopez1997} :
\be
G(x,t)=4\int_{k_0}^{\infty} \ {dk \over 2\pi} 
  \left[ 1-\cos\left(kx\right) \right] S(k,t) \ ,
\label{eq:GvsS}
\ee
where the lower cutoff $k_0$ of the integration is decided by
the system size $L$, $k_0=2\pi/L$.
Usually the lower cutoff can be replaced by $0$ and
Eq.~(\ref{eq:GvsS}) leads to $G(x,t)$ satisfying the FV
scaling. For $\alpha>1$, however, the integration has 
the infrared divergence and the lower cutoff becomes important.
In this case, by taking explicit care of the lower cutoff 
and from the FV scaling of $S(k,t)$ 
[for the FV scaling of $S(k,t)$, see, for example, Ref.~\cite{Barabasi}], 
one can obtain for $\xi(t) \gg x$,
\be
G(x,t)=|x|^{2\alpha}u(x/t^{1/z})+ 
  |x|^{2}\xi^{2(\alpha-1)}(t)v(x/\xi(t)),
\label{eq:Gscale2}
\ee
where $u(y\rightarrow 0)=u_0$ 
[$u_0 \log (1/y)$ if $\alpha$ is an integer], and 
$v(y)$ is an even function of $y$,
$v(y) =\sum_{i=0}^{\infty} v_{2i} y^{2i} $.
In Eq.~(\ref{eq:Gscale2}), the first term is the usual Family-Vicsek 
contribution and the second is the anomalous contribution coming from
the integration near the lower cutoff. 
The even form of the scaling function
$v(y)$ originates from the series expansion of the only 
$x$-dependent factor 
$1-\cos (kx)$ in Eq.~(\ref{eq:GvsS}).
Notice that the anomalous contribution is dominant over
the FV contribution for $\alpha>1$.

To demonstrate the universality, we show that
the result in Eq.~(\ref{eq:w2Nscaling}), regarding the scaling of
$w^2_N(l,t)$, comes naturally from the form of $G(x,t)$ 
in Eq.~(\ref{eq:Gscale2}).
For simplicity, let us take $N=1$. 
The generalization to a larger value of $N$ is also straightforward
and it will be sketched below.
The key element of the demonstration is to derive a general
relation between $w^2_1(l,t)$ and $G(x,t)$.
From the definition of $w^2_1(l,t)$, one can verify the following 
linear relation : 
\be
w^2_1(l,t)={1 \over 2l^2}\int_{X_l} dx_1 dx_2 \ G(x_1-x_2,t)
  \left( 1+ {12 \over l^2}x_1 x_2 \right)  ,
\label{eq:genmodw21}
\ee
where the center of the local window $X_l$ is chosen as
the origin of the coordinate system.
Then by combining Eq.~(\ref{eq:genmodw21}) with Eq.~(\ref{eq:Gscale2}),
one finds
\be
w^2_1(l,t) = \sum_{i=0}^{\infty}f_1(i+1)  v_{2i} l^{2\alpha} 
   \left( {\xi(t)\over l} \right)^{2(\alpha-1-i)} 
   +f_1(\alpha)  u_0 l^{2\alpha} + \cdots \ ,
\label{eq:w21genseries}
\ee
where
\be
f_1(\mu)={4(1-\mu) \over (2\mu+1)(2\mu+2)(2\mu+4) } \ ,
\ee
and the dots represent other terms smaller than $l^{2\alpha}$.
Then to find out the scaling behavior of $w^2_1(l,t)$, one just
needs to select the leading order term from 
Eq.~(\ref{eq:w21genseries}). In power counting of $l/\xi(t)$, 
$f_1(1)v_0 l^2 \xi(t)^{2(\alpha-1)}$ appears to be the leading order
term. However this term is not the true leading order term,
since the proportionality constant $f_1(1)$ vanishes identically.
Then by choosing the next leading order term in power counting,
one can show that Eq.~(\ref{eq:w21MBE}), which is a special case
of Eq.~(\ref{eq:w2Nscaling}), is universal. 

The generalization of the analysis to a larger value of $N$
can be carried out in a similar way.
From the definition of $w^2_N(l,t)$, one can derive a linear relation
between $w^2_N(l,t)$ and $G(x,t)$. By combining this relation
with Eq.~(\ref{eq:Gscale2}), one finds
\be
w^2_N(l,t)= \sum_{i=0}^{\infty}f_N(i+1)  v_{2i} l^{2\alpha} 
   \left( {\xi(t)\over l} \right)^{2(\alpha-1-i)}  
    +f_N(\alpha)  u_0 l^{2\alpha} + \cdots \ ,
\label{eq:w2Ngenseries}
\ee
where the dots represent other terms smaller than $l^{2\alpha}$, 
and $f_N(\mu)$ is zero if $\mu$ is an integer smaller than or
equal to $N$ \cite{comment}.
Then by selecting the true (nonvanishing) leading order term  
from Eq.~(\ref{eq:w2Ngenseries}), one can show the universality
of Eq.~(\ref{eq:w2Nscaling})

So far, we have demonstrated the universality of the macroscopic
background formation only for $1$+$1$ dimensional systems.
However the universality is not restricted to 1+1 dimensional systems.
For illustration, here we show the universality of Eq.~(\ref{eq:w21MBE})
for $D$+1 dimensional systems.
The generalization of the two key relations 
Eqs.~(\ref{eq:GvsS},\ref{eq:genmodw21}) can be achieved
through trivial replacements such as $kx \rightarrow {\bf k}\cdot{\bf x}$
and $x_1 x_2 \rightarrow {\bf x_1}\cdot {\bf x_2}$,
where a $D$-dimensional cube with volume $l^D$ is chosen 
as a $D$-dimensional local window.
Then following the same procedure, one obtains
precisely the same expression as Eq.~(\ref{eq:w21genseries})
except for a replacement of $f_1(\mu)$ by $f_{1,D}(\mu)=Df_1(\mu)$,
and one verifies the universality in higher spatial dimensions.
The generalization of the analysis to a larger value of $N$ 
is also straightforward.

One important signature of the super-rough growth models is
the universal value of the local roughness exponent $\alpha'=1$.
For some growth models \cite{Amaral,Jensen,Dong},
numerical simulations resulted in $\alpha'\approx 1$,
and so it would be interesting to examine 
whether the present analysis applies
to those growth models. 
One of the most promising candidates
is the 1+1 dimensional model of driven interfaces in random media, for
which the functional renormalization group study by
Narayan and Fisher \cite{Narayan} resulted in $\alpha'=1$ 
in the critical region.

In this paper, we consider the anomalous roughening in
super-rough $(\alpha>1)$ growth models whose structure factors
satisfy the Family-Vicsek scaling.
We show that these growth models share the following universal features. 
First, the local surface profile is characterized by 
a formation of a macroscopic background, which takes a form
of an $N$th order polynomial, where $N$ is the largest integer
smaller than the global exponent $\alpha$.
Second, after subtracting the macroscopic background,
the width of the resulting local surface profile satisfies
the Family-Vicsek scaling.
 
\vspace{0.5cm}
We would like to acknowledge helpful discussions with J. M. Kim.
This work is supported by the Korea Science and 
Engineering Foundation through the SRC program of SNU-CTP.

\begin{references}

\bibitem{Family} {\it Dynamics of Fractal Surfaces}, edited by F. Family 
  and T. Vicsek (World Scientific, Singapore, 1991), Chap. 3, p. 73.
\bibitem{Barabasi} A.-L. Barab\'{a}si and H. E. Stanley,
  in {\it Fractal Concepts in Surface Growth} 
  (Cambridge University Press, Cambridge, 1995).
\bibitem{FVansatz} F. Family and T. Vicsek, J. Phys. A {\bf 18},
  L75 (1985).
\bibitem{Amar} J. G. Amar, P.-M. Lam, and F. Family, Phys. Rev. E
  {\bf 47}, 3242 (1993).
\bibitem{Amaral} L. A. N. Amaral, A. L. Barabasi, H. A. Makse,
  and H. E. Stanley, Phys. Rev. E {\bf 52}, 4087 (1995).
\bibitem{Schroeder} M. Schroeder, M. Siegert, D. E. Wolf, J. D. Shore,
  and M. Plischke, Europhys. Lett. {\bf 24}, 563 (1993).
\bibitem{DasSarma} S. Das Sarma, S. V. Ghaisas, and J. M. Kim,
  Phys. Rev. E {\bf 49}, 122 (1994).
\bibitem{Lopez1997} J. M. L\'{o}pez, M. A. Rodr\'{i}guez,
  and R. Cuerno, Report No. cond-mat/9703024.
\bibitem{Lopez} J. M. L\'{o}pez and M. A. Rodr\'{i}guez,
  Phys. Rev. E {\bf 54}, R2189 (1996). 
\bibitem{commentAmar} The formation of the linear background
  has been previously identified by Amar {\it et al.} \cite{Amar} 
  as an origin of the anomalous roughening for the $m=2$ uniform 
  diffusion model, which they call groove instability.
\bibitem{commentN} The proportionality constant in front of
  the minimal discrepancy $l^{2\alpha}$ decays as $N\rightarrow \infty$.
  However, this decay does not change the scaling of the minimal
  discrepancy since the relevant limiting procedure is to take
  $\xi(t)/l \rightarrow \infty$ first and then allow
  $N \rightarrow \infty$.
\bibitem{pattern} For example, see M. L\"{u}cke, M. Mihelcic, 
  B. Kowalski, and K. Wingerach, in
  {\it The Physics of Structure
  Formation}, edited by W. G\"{u}ttinger and G. Dangelmayr
  (Springer-Verlag, Berlin, 1987).
\bibitem{comment} For an integer $\alpha$, the FV term 
  $f_N(\alpha)u_0 l^{2\alpha}$ in Eq.~(\protect\ref{eq:w2Ngenseries})
  is modified to 
  $f'_N(\alpha)u_0 l^{2\alpha} \log [\kappa_N(\alpha)\xi(t)/l]$
  where $\kappa_N(\alpha)$ is a constant of order 1.  
  The modification occurs because the $y\rightarrow 0$ limiting behavior
  of the scaling function $u(y)$ [Eq.~(\protect\ref{eq:Gscale2})]
  is modified from $u_0$ to $u_0 \log (1/y)$ for an integer $\alpha$. 
  The new proportionality constant $f'_N(\alpha)$ is not zero
  and so the FV term survives even if $\alpha$ is an integer.
\bibitem{Jensen} H. J. Jensen, J. Phys. A {\bf 28}, 1861 (1995).
\bibitem{Dong} M. Dong, M. C. Marchetti, A. A. Middleton, and
  V. Vinokur, Phys. Rev. Lett. {\bf 70}, 662 (1993).
\bibitem{Narayan} O. Narayan and D. S. Fisher, Phys. Rev. B
  {\bf 48}, 7030 (1993).

\end{references}
\end{document}